**Gender Inequities Throughout STEM**

**Compared to Men, Women with Significantly Higher Grades Drop STEM Majors**

*By Alexandru Maries, Kyle Whitcomb, and Chandralekha Singh*

**Abstract**

Efforts to promote equity and inclusion using evidence-based approaches are vital to correcting long-standing societal inequities that have disadvantaged women and discouraged them from pursuing studies, including in many STEM disciplines. We used 10 years of institutional data from a large public university to investigate the grade point average trends in different STEM disciplines for men and women who declared a major and then either completed the degree or dropped the major after declaring it. We found alarming trends, such as that women who dropped majors on average earned higher grades than men, and in some STEM majors, women who dropped the majors were earning comparable grades to men who persisted in those majors. While these quantitative findings call for a deeper understanding of the reasons women and men drop a major, we provide suggestions for approaches to make learning environments more equitable and inclusive so that traditionally excluded stereotyped groups can have a higher sense of belonging and thrive.

**Introduction**

Increasingly, science, technology, engineering, and mathematics (STEM) departments across the United States are focusing on using evidence to improve the learning of all students, regardless of their background; make learning environments equitable and inclusive (Watkins & Mazur, 2013; Johnson et al., 2017; Packard et al., 2020; Freeman et al., 2020; Maltese & Tai, 2011; Sevian & Robinson, 2011; Borrego & Henderson, 2014; Henderson et al., 2012); and improve their STEM retention more broadly (Kamen & Leri, 2019; Pawloski & Shabram, 2019; Callens et al., 2019). However, women are still severely underrepresented in many STEM disciplines (National Student Clearinghouse, 2015; National Science Board, 2018). To understand the successes and shortcomings of the current state of education, the use of institutional data to investigate past and current trends is crucial. In the past few decades, institutions have been keeping increasingly large digital databases of student records, and these data can be used to conduct robust statistical analyses (Baker & Inventado, 2014; Papamitsiou & Economides, 2014) that can provide invaluable information for transforming learning environments so they are more equitable and inclusive for all students (Ohland et al., 2008; Lord et al., 2009; Eris et al., 2010; Min et al., 2011; Lord et al., 2015; Ohland & Long, 2016; Matz et al., 2017; Witherspoon & Schunn, 2019; Safavian, 2019).

In this study, we use 10 years of institutional data from a large, public, doctoral university which produces a high level of research, to investigate how patterns of student major-declaration and subsequent degree-earning may differ for men and women.

Prior research has proposed a few mechanisms by which historical societal stereotypes and biases about gender may influence a student's choice of major. For example, Leslie and colleagues (Leslie et al., 2015) showed that disciplines with a higher attribution of "brilliance" also have a lower representation of women due to pervasive stereotypes about men being "brilliant" in those disciplines. These brilliance attributions affect all levels of STEM education, starting with early childhood, where girls have already acquired notions that they are not as brilliant as boys (Bian et al., 2017; Bian, Leslie, & Cimpian 2018), which can later influence their interest in pursuing certain STEM disciplines (Bian et al., 2018) and even affect how likely they are to be referred for employment in these disciplines in professional contexts (Bian, Leslie, & Cimpian, 2018).

Additionally, Eccles's expectancy-value theory (EVT; Eccles et al., 1984; Eccles et al., 1990; Eccles, 1994) can also provide a lens through which we can understand disparities in degree-earning between men and women. EVT states that a student's persistence and engagement in a discipline are related to the student's expectancy about their success, as well as how the student values the task. In an academic context, "expectancy," which refers to the individual's beliefs about their success in the discipline, is closely related to Bandura's construct of self-efficacy (sometimes referred to as competence beliefs). This construct is defined as one's belief in one's capability to succeed at a particular task or subject (Bandura, 1991, 1997, 1999, 2005), and self-efficacy has been shown to impact students' interests (Zimmerman, 2000) in pursuing studies in a discipline.

There are four main factors that influence students' expectancy or self-efficacy (Bandura, 1991, 1997, 1999, 2005; Zimmerman, 2000): vicarious experiences (e.g., instructors or peers as role models), social persuasion (e.g., explicit mentoring, guidance, and support), level of anxiety, and performance feedback (e.g., via grades on assessment tasks). Women generally have lower self-efficacy than men in many STEM disciplines because these four factors can negatively influence them (Johnson et al., 2017; Tolbert et al., 2018; Marshman et al., 2018; Bang & Medin, 2010; Estrada et al., 2018; Ong et al., 2018). For example, in many STEM fields, women are underrepresented in their classrooms and less likely to have a female role model among the faculty (National Student Clearinghouse, 2015; National Science Board, 2018). Furthermore, the stereotypes surrounding women in many STEM disciplines can affect how they are treated by mentors, even if such an effect is subconscious (Marchand & Taasoobshirazi 2013; Moss-Racusin et al., 2012; Hill et al., 2010; Appel & Kronenberger, 2012; Wheeler & Petty, 2001). Moreover, women are susceptible to stress and anxiety from stereotype threat (i.e., the fear of confirming stereotypes about women in many STEM disciplines), which their male peers are less likely to experience. This stress and anxiety can rob them of their cognitive resources (Beilock et al., 2006; Beilock et al., 2007), especially during high-stakes assessments such as exams.

Expectancy can influence grades earned as well as the likelihood that a student will persist in a program (Eccles et al., 1984; Eccles et al., 1990; Eccles, 1994). Stereotype threats that women in many STEM disciplines experience can increase anxiety in learning and test-taking situations and lead to deteriorating performance. Because anxiety can increase when performance deteriorates, these factors working against women in STEM can force them into a feedback loop and hinder their performance further, which can lower their self-efficacy even more and continue to affect future performance (Bandura, 1991, 1997, 1999, 2005; Zimmerman, 2000). Moreover, students with lower self-efficacy are more likely to leave a discipline despite performing adequately.

In EVT, value is typically defined as having four facets: intrinsic value (i.e., interest in the task), attainment value (i.e., the importance of the task for the student's identity), utility value (i.e., the value of the task for future goals such as career), and cost (i.e., opportunity cost or psychological effects such as stress and anxiety; Eccles et al., 1984; Eccles et al., 1990; Eccles, 1994). In the context of women's enrollment and persistence in many STEM disciplines, societal stereotypes can influence all facets of the students' value of these STEM disciplines. Intrinsic value can be informed by societal stereotypes and brilliance attributions of the STEM disciplines, and attainment and utility values can be further tempered by these stereotypes. Utility value is an important facet of student education in STEM, since a degree in a STEM field provides many job opportunities for graduating students. In addition, the psychological cost of majoring in these disciplines can be inflated by the stereotype threat. All of these effects can conspire to suppress the likelihood of women choosing and persisting in various STEM disciplines.

To measure the long-term effects of these systemic disadvantages, we investigate differences between male and female students who persist in a STEM major and those who switch out of STEM majors. In particular, we investigate how the grade point averages (GPAs), both in STEM courses only and overall, differ between students who persist in a STEM major and those who do not. We focus on the disciplines shown in Table 1.

**Table 1**

Majors Considered in This Study and the Shortened Labels Used in Tables and Figures

| Major | Short label |
| --- | --- |
| Biological sciences and neuroscience | Bio |
| Computer science | CS |
| Engineering | Engr |
| Mathematics and statistics | Math |
| Chemistry | Chem |
| Physics and astronomy | Phys |
| Geology and environmental science | Geo |

*Note.* "Engineering" is a combination of many engineering majors offered by the School of Engineering.

The university provided for analysis the records for 18,319 undergraduate students enrolled in majors from the programs listed in Table 1. This sample of students was 49.9% female and had the following racial and ethnic backgrounds: 77.7% White, 11.1% Asian, 6.8% Black, 2.5% Hispanic, and 2.0% other or multiracial.

Measures of student academic performance were also included in the provided data in the form of grade points earned by students in each course taken at the university. Grade points are on a 0-4 scale with A = 4, B = 3, C = 2, D = 1, and F = 0, where the suffixes "+" and "−" add or subtract, respectively, 0.25 grade points (e.g., $B- = 2.75$), with the exception of A+, which is reported as the maximum of 4 grade points. These data were used to calculate students' GPA across courses taken in each year of study, as well as across STEM courses only, from their first to sixth years. STEM courses were considered to be those courses taken from the programs listed in Table 1. For the purposes of this paper, "STEM" does not include the social sciences.

The data also included the year in which the students who started in a fall semester took each course, with courses taken over the summer omitted. For example, if a student first enrolled in fall 2007, then their "first year" occurred during fall 2007 and spring 2008, their "second year" during fall 2008 and spring 2009, and so on.

**Students who drop out of a STEM major have lower GPAs than students who persist**

Figure 1 plots the mean GPA of STEM students who declared different sets of majors and then either earned a degree within that set of majors or dropped those majors. The majority of students who dropped a STEM major either did not get a degree at all or got a degree in a non-STEM discipline. There were a few cases where students switched from one STEM major to another: For students who dropped physics, roughly 16% switched to math (these students were counted in the same category of math and physical science in Figure 1), and 11% switched to engineering. There were only two other instances in which more than 10% of students who dropped a STEM major switched into a different STEM major: 16% of students who dropped chemistry switched to biology, and 12% of students who dropped geology switched to biology. Both overall GPA (Figures 1a, 1c, and 1e) and STEM GPA (Figures 1b, 1d, and 1f) are plotted. Across all of Figure 1, the large drop in sample size from year 4 to 5 and again from 5 to 6 is primarily due to students graduating. While the data for 6 years are plotted for completeness, we will focus our analysis on years 1 through 4. Note that the number of students for a particular year in the overall GPA are sometimes not exactly the same as the number in the same year for the STEM GPA because there were a few students in those years who did not take STEM courses.

Figure 1 shows that in general, the students who dropped any given major have a lower GPA and STEM GPA than students who earned a degree in that major. However, the difference between the two groups varies based on which cluster of majors we consider. For biological science and neuroscience majors (Figures 1a and 1b), as well as mathematics and physical science majors (Figures 1e and 1f), those who persisted in the major have a GPA of roughly 0.3 to 0.6 grade points higher than those who dropped, with similar differences in STEM GPA. This difference in grade points represents a difference of one to two letter grades at the studied university, where, for example, the difference between a B and B+ is 0.25 grade points and between a B+ and A is 0.5 grade points. For computer science and engineering majors (Figures 1c and 1d), those who persisted in the major have a GPA of roughly 0.6 points higher than those who dropped.

**Performance difference between women who drop and men who drop is larger than performance difference between women who persist and men who persist**

Figure 2 shows the same data as Figure 1, except separated by gender. In all cases, the main finding is that women are earning higher grades on average than men, both among those who drop a given major and those who earn a degree in that major. However, the grade differences between men and women who dropped the major and men and women who earned a degree in the major are inconsistent. Across all majors in Figure 2, the differences between men and women who dropped a given major are larger than the differences between men and women who earned a degree in that major. This is particularly noticeable in Figures 2c, 2d, 2e, and 2f, where by the fourth year the women who had dropped these majors (computer science, engineering, mathematics, and physical science) were earning nearly the same GPA and STEM GPA as the men who persisted in these majors.

It is also noteworthy that in certain cases, the GPAs of women who dropped a major in a given year were very similar to, and sometimes indistinguishable from, the GPAs of men who persisted. For example, for computer science and engineering majors, the difference in overall GPA between women who dropped and men who persisted in years 3 and 4 is only roughly 0.2 and 0.1 grade points, respectively, and these differences are much smaller than the differences between women who dropped and men who dropped (roughly 0.5 grade points for both years). A similar result occurs for year 4 in math and physical science: There was a difference in overall GPA of 0.2 grade points between women who dropped and men who persisted, but a difference of 0.5 between women who dropped and men who dropped. It is possible that women who dropped these majors in these years did not persist in the major due to inequitable and non-inclusive learning environments and lack of positive recognition from instructors about how they were doing.

**Figure 1.** Mean Overall GPA and STEM GPA, Along With Standard Error Over Time for STEM Majors

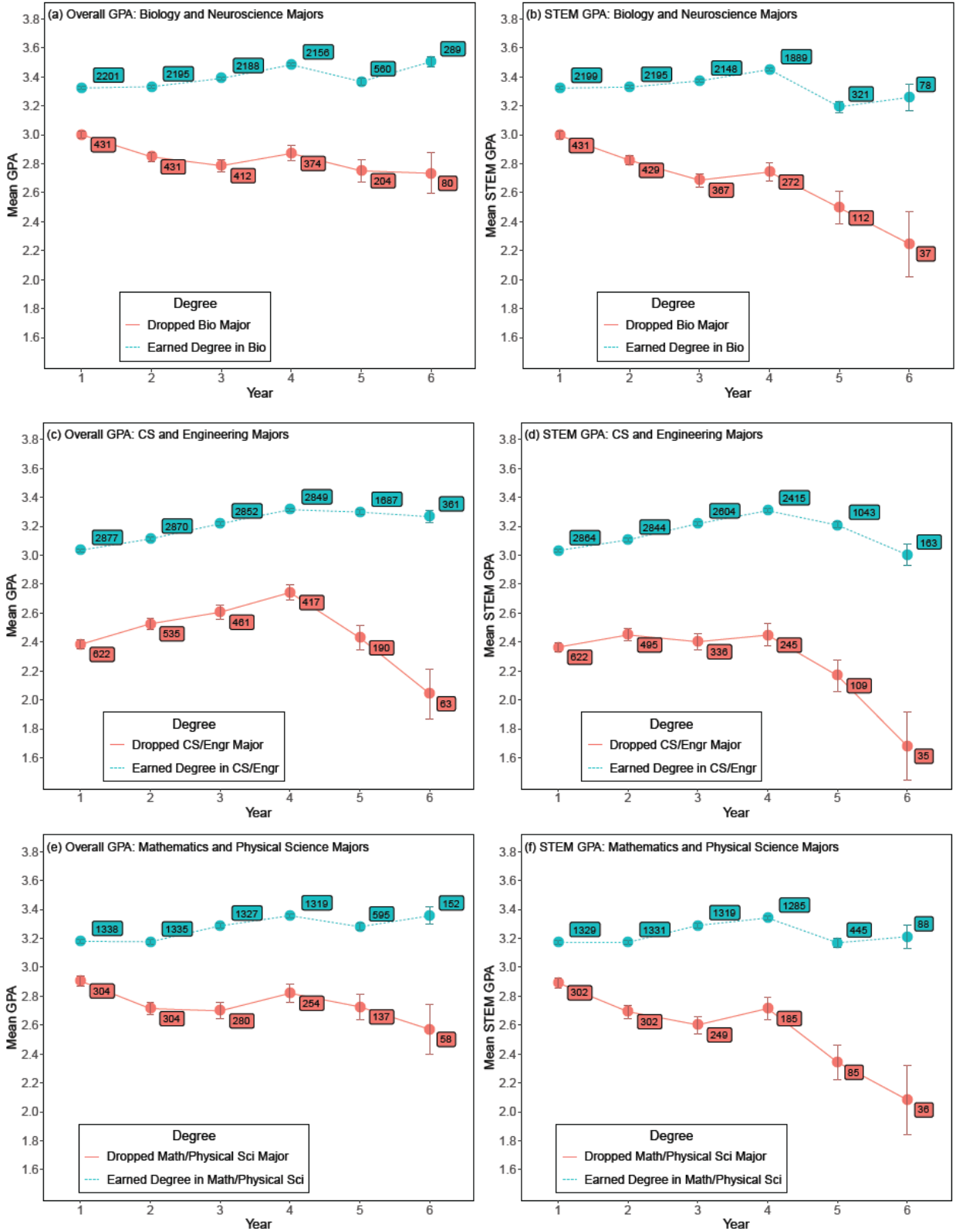

*Notes.* Mean overall GPAs are shown in 1a, 1c, and 1e; mean STEM GPAs are shown in 1b, 1d, and 1f. Sample size is listed by each point and lines connecting the points are included to help identify trends. Majors listed in plot legend entries.

**Figure 2.** Mean Overall GPA and STEM GPA, Along With Standard Error Over Time for STEM Majors by Gender

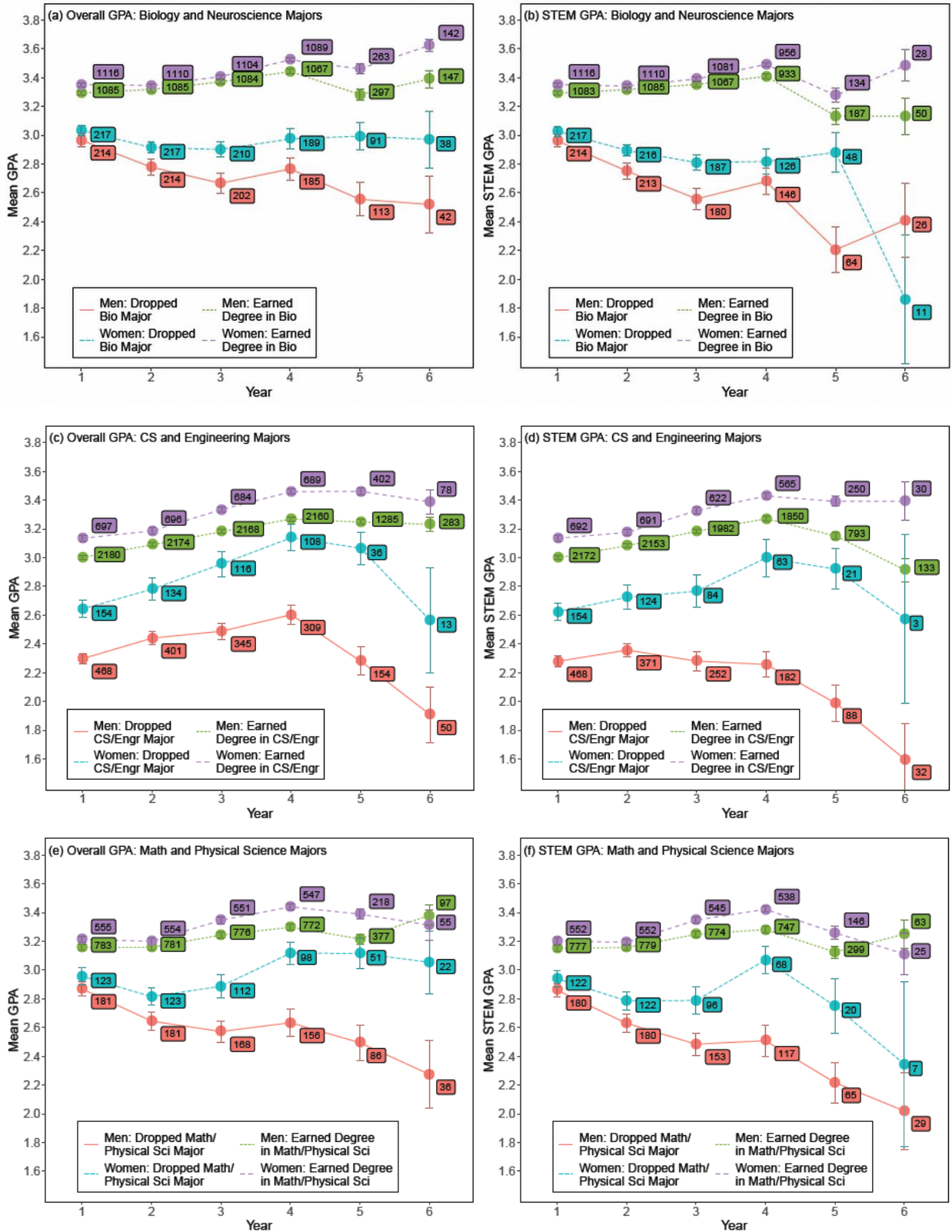

*Notes.* Mean overall GPAs are shown in 1a, 1c, and 1e; mean STEM GPAs are shown in 1b, 1d, and 1f. Sample size listed by each point and lines connecting the points are included to help identify trends.

To compare men and women from both groups (dropped and persisted), we report in Tables 2a and 2b the results from four comparisons. In Table 2a, we compare the STEM GPA of the women who dropped each set of STEM majors (Figures 2b, 2d, and 2f) with the STEM GPA of men who dropped those same majors, and in Table 2b, we compare the STEM GPA of women who earned a degree in the STEM majors with men who earned a degree in the same majors. These comparisons are performed with two measures: the *p* value from a two-tailed *t*-test (Cohen, 1988) and the effect size from Cohen's *d* (Cohen, 1988).

In each set of majors, we find that the mean STEM GPA difference between women and men who dropped each set of STEM majors is larger than the STEM GPA difference between women and men who persisted in the majors, and sometimes (in the case of computer science and engineering majors), the difference between women and men who dropped the majors is medium. In both cases, the women were earning higher grades than the men. Thus, the women who dropped these majors had, on average, better grades than their male counterparts by one or two letter grades, while the women who persisted earned similar grades to the men who persisted. This indicates that women are more likely to drop the major than men with the same grades when those grades fall below the mean.

TABLE 2. A Comparison of the Cumulative GPA Between Men and Women Who (a) Drop the Same STEM Majors and (b) Earn a Degree in the Same STEM Major

| (a) | Women who dropped the major | | | Men who dropped the major | | | Statistical comparisons | |
|---|---|---|---|---|---|---|---|---|
| Figure | N | M | SD | N | M | SD | p | d |
| 2b | 217 | 2.95 | 0.52 | 214 | 2.76 | 0.73 | .002 | 0.30 |
| 2d | 154 | 2.68 | 0.78 | 468 | 2.23 | 0.81 | < .001 | 0.56 |
| 2f | 123 | 2.91 | 0.64 | 181 | 2.66 | 0.74 | < .001 | 0.36 |

| (b) | Women who earned a degree | | | Men who earned a degree | | | Statistical comparisons | |
|---|---|---|---|---|---|---|---|---|
| Figure | N | M | SD | N | M | SD | p | d |
| 2b | 1116 | 3.40 | 0.40 | 1085 | 3.36 | 0.44 | .017 | 0.10 |
| 2d | 696 | 3.26 | 0.46 | 2180 | 3.13 | 0.52 | < .001 | 0.27 |
| 2f | 555 | 3.30 | 0.44 | 783 | 3.23 | 0.50 | < .001 | 0.15 |

*Note:* The numbers (*N*) refer to the total number of unique students in each group (dropped the major or earned a degree). The averages (*M*) refer to the average cumulative STEM GPA of all students in that group at the time when they either (a) graduated with a degree or (b) dropped the major, and SD refers to the standard deviation of the cumulative STEM GPA for all students in each group. The p-value from a two-tailed *t*-test is reported comparing the women and men, along with Cohen's d measuring the effect size of the gender difference (the sign of d matches the sign of the mean GPA for women minus the mean GPA for men). The comparison is performed separately for the STEM majors corresponding to the indicated figure, i.e., for Fig. 2b we consider biological science and neuroscience majors; for Fig. 2d we consider computer science and engineering majors; and for Fig. 2f we consider mathematics and physical science majors.

**Discussion and implications**

We find pervasive and deeply troubling gendered trends in the overall GPA and STEM GPA of students who drop different STEM majors. In particular, Figure 2 shows that the women who dropped a STEM major have a higher average GPA than the men who dropped a STEM major, and these differences are shown to be statistically significant in Table 2a. This is in contrast with comparisons of men and women who persisted and earned degrees in STEM majors, where women still had higher grades, but with only a small effect size, as shown in Table 2b. Thus, on average, among students with the same GPA, the women in all STEM majors were more likely than the men to drop the majors, while the men were more likely to earn a degree. It is important to note that this is true among the STEM majors with gender imbalances in the population as well as those majors without imbalances.

While it is true that women at the studied university on average earned higher grades than men in most STEM courses, that does not explain why women chose to drop STEM majors with the same grades as men who persisted. The difference must then be coming from another source, such as inequitable and non-inclusive learning environments and lack of mentoring and support for women who may have lower self-efficacy (Bandura, 1991, 1997, 1999, 2005; Zimmerman, 2000). Women may also have a lower sense of belonging and value pertaining to remaining in these disciplines (Safavian, 2019; Eccles et al, 1984; Eccles et al., 1990, Eccles, 1994) if learning environments are not equitable and inclusive, especially because they must bear the high cost of managing the burden of societal stereotypes and the ensuing stereotype threats. Prior research suggests that the brilliance attributions of STEM disciplines (Dweck, 2006; Leslie et al., 2015) and who is likely to excel in these disciplines could influence women to move away from a discipline in which they could have succeeded. Furthermore, lack of recognition has been shown to negatively impact the interest and self-efficacy of women in physics (Kalender et al., 2019); since both of these motivational constructs have been shown to relate to persistence, lack of recognition is also likely to negatively impact whether a woman persists in a STEM discipline. Thus, without explicit efforts by faculty to improve the learning environments in these disciplines, the culture and stereotypes surrounding STEM in general may create an environment in which women are being unfairly driven out of these fields in which they could have thrived, while their male counterparts are not subjected to these same pressures and persist in their studies despite worse performance.

Given all this, what can instructors do to help improve the persistence of all students, but especially women? Prior research has suggested several promising avenues that focus on creating an equitable and inclusive learning environment.

*Address motivational factors of self-efficacy and social belonging, and promote a growth mindset.*

The literature on psychological interventions has shown that explicitly addressing self-efficacy by promoting a growth mindset and incorporating activities that help provide narratives to scaffold students' experiences with adversity can have a long-lasting beneficial impact that reduces achievement gaps (Yeager & Dweck, 2012; Binning et al., 2020). These interventions can be used at the beginning of the term, but the culture of the class must be consonant with the message of the interventions (Canning et al., 2019).

*Help students interpret their performance accurately*.

Inaccurate self-assessments of self-efficacy can negatively impact students' motivation and interest, and students with higher self-efficacy have been shown to engage in better self-regulated learning strategies (Zimmerman, 2000). Faculty can help students interpret their grades (e.g., a B is above average), but some students (more likely women) may interpret it as a poor grade. Additionally, providing micro-affirmations during group work to help build students' confidence can be beneficial. Training teaching assistants (TAs) is also important, because they often have more chances to interact with students in smaller recitation or lab sections. For example, in an interview regarding self-efficacy, a woman in an introductory physics class (Kalender et al., 2019) described how when she asked her first question in a recitation, the TA replied, "Oh, that is an easy question," making her feel embarrassed; consequently, she never asked another question during recitation. The TA may not have meant anything negative by that comment, but it is important that TAs recognize that nothing is easy for students who are novices learning something new, and therefore these students will struggle with things that can seem simple to the TA.

*Provide recognition for students' efforts or improvement.*

A recent study of motivational factors that relate to interest in physics (Kalender et al. 2019) found that perceived recognition from others has a strong influence on both self-efficacy and interest, which both relate to persistence. Students are often working hard in their coursework, and micro-affirmations when they are doing well can go a long way. One should never assume that the quality of students' work is self-evident, as shown in Eileen Pollack's story (2013) about working on her undergraduate thesis. Pollack had worked with a mathematics professor on an independent project during which she had to solve a challenging problem. After spending many months working on the problem, she eventually figured it out, but when she triumphantly went to her adviser's office, she did not receive any recognition or encouragement, prompting her to drop out of mathematics and pursue literature. Years later, when she met with her adviser, she asked him about this work, and he said that given how unusual it is for an undergraduate student to do an independent project, he would characterize the work as exceptional. Had he told her that when she completed this work, she may have chosen to pursue a mathematics career. The same is so often true for introductory students in STEM who are most likely to switch a major during their first year. Providing encouragement that helps students develop a sense of self-efficacy can help them persist at such a critical time (Mervis, 2010).

In conclusion, from analyzing 10 years of institutional data, we find alarming trends of women who drop out of a STEM major having significantly higher GPAs than men who drop those same majors, and sometimes, having GPAs that are indistinguishable from men who persist in those majors. This may be due to inequitable learning environments in these STEM courses and instructors can do a lot (in addition to making use of evidence-based active learning strategies) to support the learning of all students and reduce gender disparities, such as address the motivational factors of self-efficacy, social belonging, and growth mindset, help students interpret performance accurately, and provide recognition for students' efforts or improvement.